\documentclass[a4paper,11pt]{article}
\usepackage{pos}
\usepackage{amsmath}
\usepackage{braket}
\usepackage{macros_AS} 
\usepackage{slashed,comment,bbm}

\title{Probing higher moments of pion parton distribution functions}
\ShortTitle{Flowed moments}

\author[a]{Anthony~Francis}
\author[b]{Patrick~Fritzsch}
\author[c,d]{Rohith~Karur}
\author[e]{Jangho~Kim}
\author[f,g]{Giovanni~Pederiva}
\author*[c,d]{Dimitra~A.~Pefkou}
\author[h]{Antonio~Rago}
\author*[i,d,c]{Andrea~Shindler}
\author[d,c]{Andr\'e~Walker-Loud}
\author[j]{Savvas~Zafeiropoulos}

\affiliation[a]{Institute of Physics, National Yang Ming Chiao Tung University, 30010 Hsinchu, Taiwan}
\affiliation[b]{School of Mathematics, Trinity College, College Green, Dublin, Ireland}
\affiliation[c]{Department of Physics, University of California, Berkeley, CA 94720, U.S.A}
\affiliation[d]{Nuclear Science Division, Lawrence Berkeley National Laboratory, Berkeley, CA 94720, USA}
\affiliation[e]{Facult\"at f\"ur Physik, Universit\"at Bielefeld,
  Universit\"atsstra{\ss}e 25, 33615, Bielefeld, Germany}
\affiliation[f]{J\"ulich Supercomputing Center, Forschungszentrum Jülich, Wilhelm-Johnen-Straße, 54245 Jülich, Germany}
\affiliation[g]{Center for Advanced Simulation and Analytics (CASA), Forschungszentrum Jülich, Wilhelm-Johnen-Straße, 54245 Jülich, Germany}
\affiliation[h]{$\hbar$QTC \& IMADA, University of Southern Denmark, Campusvej 55, 5230 Odense M, Denmark}
\affiliation[i]{Institute for Theoretical Particle Physics and Cosmology, TTK, RWTH Aachen University, Sommerfeldstr. 16, Aachen, 52074, Germany}
\affiliation[j]{Aix Marseille Univ, Universit\'e de Toulon, CNRS, CPT, Marseille, France}

\emailAdd{dpefkou@berkeley.edu}
\emailAdd{shindler@physik.rwth-aachen.de}

\abstract{We present the first numerical investigation of the method proposed in Ref.~\cite{Shindler:2023xpd} to utilize gradient flow to obtain precise determinations of higher moments of PDFs from lattice QCD, circumventing power divergent mixing with lower dimensional operators. We apply this method to obtain moments of the isovector PDF of the pion using four Stabilized Wilson Fermion ensembles with $m_{\pi}\simeq 411~\text{MeV}$ and lattice spacings $a\simeq 0.064, 0.077, 0.094$, and $0.12$ fm. We present preliminary results of ratios of three-point functions as a function of flow time, which can be used to extract the ratios $\braket{x^2}/\braket{x}$ and $\braket{x^3}/\braket{x}$. 
We find that a significantly higher precision can be achieved with this method compared to the canonical approach, which requires boosting and cannot reach higher than the $\braket{x^3}$ moment.}

\FullConference{The 41st International Symposium on Lattice Field Theory (LATTICE2024)\\
 28 July - 3 August 2024\\
Liverpool, UK\\}

\begin{document}
\begin{flushright}
  TTK-24-38
\end{flushright}
\maketitle

\section{Introduction}
\label{sec:intro}

The study of parton distribution functions (PDFs) is essential for interpreting data from both current and future high-energy physics experiments. PDFs describe the internal structure of hadrons in terms of their constituent quarks and gluons, providing crucial input for the precise analysis of results from particle colliders like the Large Hadron Collider (LHC) and forthcoming facilities such as the Electron-Ion Collider (EIC).

Lattice QCD provides a first-principles approach to studying hadron structure, offering a method to directly compute PDFs. Over the past decade, numerous approaches have been proposed to achieve direct determinations of PDFs from lattice QCD simulations\footnote{See Ref.~\cite{Cichy:2018mum} for a review and a comprehensive list of references.}. Among these, alternative methods, such as those presented in Refs.~\cite{Karpie:2018zaz,Joo:2019bzr,Gao:2022iex}, leverage non-local Wilson line operators and derivatives of the Ioffe time distribution~\cite{Orginos:2017kos}, showing potential for calculating higher moments of PDFs on the lattice.

Pioneering studies~\cite{Kronfeld:1984zv,Martinelli:1987zd,Martinelli:1987bh} have already identified the limitations of this approach. The breaking of continuous symmetries at finite lattice spacing forces quantum corrections to respect the residual hypercubic symmetry H(4). As a result, the renormalization of local fields, related to PDF moments, is complicated by mixings with lower-dimensional fields, which induce power divergences with the lattice spacing, $a$. For moments up to $\left\langle x^3 \right\rangle$, this mixing can be avoided by introducing a non-zero spatial momentum, though this increases the noise-to-signal ratio in correlation functions. However, for higher moments $\left\langle x^n \right\rangle$ with $n>3$, power divergences become unavoidable, effectively preventing the determination of higher moments.

In Ref.~\cite{Shindler:2023xpd}, one of the co-authors of this contribution proposed a novel methodology to address the challenge of determining moments of any order of PDFs, while simultaneously improving the noise-to-signal ratio for moments that can be approached with standard methods. This approach utilizes the gradient flow (GF) for gauge~\cite{Narayanan:2006rf,Luscher:2010iy,Luscher:2011bx} and fermion~\cite{Luscher:2013cpa} fields as an intermediate regulator to restore the continuum O($4$) symmetry before matching the results obtained at finite flow time, $t$, using perturbation theory.

In this proceeding, after a brief recap of the method described in~\cite{Shindler:2023xpd}, we present preliminary results for the moments of pion PDFs in the non-singlet case. These preliminary results confirm expectations, allowing the determination of moments up to \(\left\langle x^5 \right\rangle\) with moderate computational effort, while simultaneously improving the noise-to-signal ratio compared to standard techniques.

\section{Flowed moments}
\label{sec:flowmom}

Moments of PDFs are related to hadronic matrix elements of local operators.
In the non-singlet case, the dominant contribution at large $4$-momentum transfer to the nucleon, $Q^2=-q^2$, as seen in DIS experiments, comes from twist-2 operators
\be 
O_n^{rs}(x) = O^{rs}_{\left\{\mu_1 \cdots \mu_n\right\}}(x) =
\psibar^r(x) \gmuopen1 \lrDmu2 \cdots \lrDmuclose{n} \psi^s(x)\,,
\label{eq:t2}
\ee 
where $\psi$ and $\psibar$ are fermion fields, $\lrD_{\mu_i} = 1/2\left(\rD_{\mu_i} - \lD_{\mu_i}\right)$ denotes symmetrized covariant derivatives and 
$\left\{ \mu_1 \cdots \mu_n\right\}$ indicates normalized symmetrization over the Lorentz indices
\bea
O^{rs}_{\{\mu_1...\mu_n\}}=\frac{1}{n!}\sum_{\sigma\in\text{permutations}}O^{rs}_{\mu_{\sigma(1)}...\mu_{\sigma(n)}}\,.
\eea
To avoid complications with mixing with gluonic operators, we focus on the non-singlet case $r \neq s$.
For unpolarized targets, these operators are directly related to the moments of PDFs of the hadron $h$ 
\be 
\left \langle h(p)| \widehat{O}_n |h(p) \right\rangle = 2
\left( p_{\mu_1} \cdots p_{\mu_n} - \text{trace~terms}\right) \left\langle x^{n-1} \right\rangle_h(\mu) \,,
\label{eq:matrix}
\ee 
where we omit the flavor indices for simplicity, and $\widehat{O}_n$ denotes 
the twist-2 operators with trace terms subtracted. 
The energy scale $\mu$ represents the renormalization scale of the local operators~\eqref{eq:t2}.

The flowed operators are defined as
\be 
O_n^{rs}(x,t) = \chibar^r(x,t) \gmuopen1 \lrDmu2 \cdots \lrDmuclose{n} \chi^s(x,t)\,,
\label{eq:flowed_t2}
\ee 
where $\chi^r(x,t)$, $\chibar^r(x,t)$, and $B_\mu(x,t)$ denote respectively fermion and gauge fields satisfying the GF equations from Refs.~\cite{Luscher:2010iy,Luscher:2013cpa}. These operators renormalize multiplicatively, with a common renormalization factor depending only on the fermion content. For the flowed fermion fields, a commonly adopted scheme introduces {\it ringed fields}~\cite{Makino:2014taa}, $\rchi$ and $\rchibar$, defined by the gauge-invariant condition
\be 
\left \langle \rchibar(x,t) {\overset\leftrightarrow{\slashed{D}}} \rchi(x,t) \right\rangle
= - \frac{N_c}{(4 \pi)^2 t^2}\,.
\ee 
This condition is particularly advantageous, as it is regularization independent and can be adopted both in perturbation theory and non-perturbatively in lattice QCD simulations.

After taking the continuum limit of hadronic matrix elements at a fixed flow time $t > 0$, the physical renormalized matrix element at $t = 0$ can be determined using the short flow time expansion (SFTX)~\cite{Luscher:2013vga}. Generally, the SFTX for twist-2 operators $O_n(x,t)$ includes power-divergent terms, where the divergence depends on the dimensions of the operator and those of the lower-dimensional operators with the same symmetry properties. These terms, however, are classified according to continuum O($4$) symmetry. Traceless operators with symmetrized Lorentz indices belong to an irreducible representation of O($4$), and as a result, their SFTX reads
\be 
\widehat{\rO}_n^{rs}(t) = c_n(t,\mu)\widehat{O}_{n,\MS}^{rs}(\mu) + \cdots\,,
\label{eq:sftx}
\ee 
where contributions at $t = 0$ come exclusively from the corresponding traceless operators, which are directly related to the moments of PDFs. The operator $\widehat{\rO}_n^{rs}(t)$ is defined as Eq,~\eqref{eq:flowed_t2}, using ringed fermion fields, and the operator at $t=0$ is renormalized in the same scheme adopted for the calculation of the matching coefficients, $c_n(t,\mu)$.

The matching coefficients are calculable using standard techniques~\cite{Manohar:2018aog,Mereghetti:2021nkt,Crosas:2023anw} and details on this calculation can be found in Ref.~\cite{Shindler:2023xpd}.
The result at next-to-leading-order (NLO) reads 
\be 
c_n(t,\mu) =  1 + \frac{\gbar^2(\mu)}{\left(4 \pi\right)^2}c_n^{(1)}(t,\mu) + 
O(\gbar^4)\,.
\label{eq:cn_1l}
\ee 
where $\gbar^2(\mu)$ is the renormalized strong coupling in the $\MS$ scheme.
The one-loop coefficient is given by
\be 
c_n^{(1)}(t,\mu) = C_F \left[ \gamma_n \log \left(8 \pi \mu^2 t \right) + B_n\right]\,,
\label{eq:matching_1l}
\ee 
where $\gamma_n = 1 + 4 \sum_{j=2}^n \frac{1}{j} - \frac{2}{n(n+1)}$, 
and provides a welcome check because it agrees with the 1-loop anomalous dimension of the twist-2 operators~\cite{Gross:1974cs}. The finite part is given by 
\bea 
B_n &=& \frac{4}{n(n+1)} + 4 \frac{n-1}{n}\log 2 + \frac{2-4 n^2}{n(n+1)}\gamma_E - \frac{2}{n(n+1)}\psi(n+2) + \frac{4}{n}\psi(n+1) - 4 \psi(2) \nonumber \\
&-& 4 \sum_{j=2}^n \frac{1}{j(j-1)} \frac{1}{2^j} \phi(1/2,1,j) - \log \left(432\right)\,,
\label{eq:finite}
\eea 
where $\phi(z,s,a)$ is the Lerch transcendent.
For generic $n$ Eqs.~\eqref{eq:matching_1l} and \eqref{eq:finite} have been determined in~\cite{Shindler:2023xpd} and they agree with the result for $n=2$ derived in~\cite{Makino:2014taa}.
The expression for $B_n$ can be written in a slightly more compact way using the relations of the digamma functions $\psi(z)$ with the harmonic numbers $H_n$
\be   
B_n = \frac{2-4n^2(n+2)}{n(n+1)^2} + \frac{2(2n+1)}{n(n+1)}H_n - \frac{4}{n} \log(2) - 3 \log(3) - 4 \sum_{j=2}^n \frac{1}{j(j-1)} \frac{1}{2^j} \phi(1/2,1,j)\,,
\ee
and perhaps further simplified using generalized harmonic sums defined in~\cite{Ablinger:2013cf}.
The same matching coefficients in the $\MSbar$ scheme are obtained with the substitution $\mu^2\rightarrow \mu^2 e^{\gamma_E}/4\pi$.
The extension to next-to-next-to-leading order (NNLO) in the perturbative calculation appears achievable with the advanced techniques currently available~\cite{Artz:2019bpr}.

The strategy for the calculation of a generic moment can be summarized in the following steps: $1)$ 
construct the symmetric and traceless operator $\widehat{O}_n^{rs}$ for a given $n$; 2) compute the hadronic matrix element $\left \langle h(p)| \widehat{\rO}_n^{rs} |h(p) \right\rangle$; 3) take the continuum limit at fixed flow time $t$, in physical units; 4) calculate the moment renormalized in the $\MS$ scheme 
\be 
\left\langle x^{n-1}\right\rangle^{\MS}(\mu) = 
c_n(t,\mu)^{-1} \left\langle x^{n-1}\right\rangle(t) \,.
\label{eq:x_MS}
\ee 
The hadronic matrix element can be calculated using a spectral decomposition of 3-point correlation functions, as described in detail in Sec.~\ref{sec:numerical}.
Since the matching in Eq.~\eqref{eq:x_MS} holds for any $n$, we have the freedom to choose the Lorentz indices of the local field. For instance, the linear combination 
\be 
O_{4444} - \sum_{\alpha=1}^4\frac{3}{4}O_{\left\{\alpha \alpha 4 4 \right\}} 
+ \frac{1}{16} \sum_{\alpha,\beta=1}^4O_{\left\{\alpha \alpha \beta \beta \right\}}\,.
\label{eq:On4}
\ee 
enables the determination of $\left\langle x^3 \right\rangle$ without requiring any spatial external momentum, thus improving the noise-to-signal ratio. This is confirmed from our numerical experiment, as discussed in Sec.~\ref{sec:results}. The matching coefficients in Eq.~\eqref{eq:cn_1l} are calculated with flowed fermions defined in a ringed scheme. Alternatively, ratios of flowed correlators with the same number of fermion fields can be analyzed. For example,
using the second moment, $\left\langle x \right\rangle_{\MS}$, as an input observable, we can determine all the other moments with 
\be 
\left\langle x^{n-1} \right\rangle^{\MS}_h(\mu) = 
\left\langle x \right\rangle^{\MS}_h(\mu)
\frac{c_{2}(t,\mu)}{c_n(t,\mu)} R_n^h(t)\,,
\label{eq:ratio_n2}
\ee 
where the ratios 
\be 
R_n^h(t) = \frac{\left\langle x^{n-1} \right\rangle_{h}(t)}{\left\langle x\right\rangle_{h}(t)}\,,
\qquad n>2\,,
\label{eq:Rn}
\ee 
are computed with bare fields and have a finite continuum limit.
The determination of ratios like the ones in Eq.~\eqref{eq:Rn} offers additional advantages: reduced statistical uncertainty, taking advantage of correlations among data, and the cancellation of cutoff effects, as discussed in Sec.~\ref{ssec:cutoff}.
The method is general and can be applied with any lattice action. With appropriate modifications, it can be also used to study distribution amplitudes and other distribution functions.

\subsection{Discretization uncertainties}
\label{ssec:cutoff}

For Wilson-type quarks, such as non-perturbatively improved clover fermions, two types of O($a$) cutoff effects remain unaccounted for beyond the usual improvement terms. First, there are O($am$) terms associated with flowed fermion fields, which are universal for any flowed fermion field and depend only on their fermion content. However, in the ratios of correlation functions used to extract the reduced matrix elements in Eq.~\eqref{eq:Rn}, these O($am$) terms cancel out.

Second, short-distance O($a$) cutoff effects may arise when the physical distance between the flowed fermion field and the unflowed hadron interpolators is small. Since extracting the reduced matrix elements~\eqref{eq:Rn} requires a large physical separation between the source (and sink) and the flowed operator, we can ensure that these O($a$) cutoff effects are negligible\footnote{A numerical investigation is currently ongoing.}. A numerical confirmation of this fact for a different correlation function can be found in Ref.~\cite{Kim:2021qae}.

One of the key advantages of this method is that specific O($a$) cutoff effects for twist-2 operators are absent, independent of $n$. As a result, the scaling toward the continuum limit for flowed moments computed with Wilson-type fermions is greatly improved compared to standard methods, where $n$-dependent O($a$) terms with unknown improvement coefficients are typically required.

\section{Numerical implementation}
\label{sec:numerical}

To test the method just described we compute the first few moments of the pion PDFs on OpenLat gauge configurations~\cite{Cuteri:2022erk,Cuteri:2022oms}. Moments of PDFs are related to the connected 3-point function, projected to vanishing spatial momentum, and given by 
\be
    C_n^{3\text{-pt}}(x_4,y_4; t) = a^6\sum_{\bx,\by} \braket{P^{du}(\bx,x_4)\widehat{\mathring{O}}_n(\by,y_4;t) P^{ud}(0) }_c \;,
    \label{eq:3ptfull}
\ee
where the interpolator $P^{ud}(x)=\overline{\psi}_u(x)\gamma_5\psi_d(x)$ has the quantum numbers of a $\pi^+$ state.
In Eq.~\eqref{eq:3ptfull} we fix the source to be at the origin to simplify the notation, but in practice we use a single $\mathbbm{Z}_4$ stochastic source per gauge configuration, with source locations randomly sampled from a uniform $4$-d probability distribution. The final result, because of translation invariance, depends only on the source-sink separation.

For $x_4\gg y_4 \gg 0$ the ground state $\ket{\pi(\bzero)}$ dominates the spectral decomposition of Eq.~\eqref{eq:3ptfull}
\be
C_n^{3\text{-pt}}(x_4; t) = \frac{|Z_\pi|^2}{4m_\pi^2}e^{-m_\pi x_4} \mathring{A}_n(t) + \cdots\,, \qquad  \mathring{A}_n(t)\equiv \bra{\pi(\bzero)} \widehat{\mathring{O}}_n(0;t) \ket{\pi(\bzero)} \;,
\label{eq:sprectral}
\ee
where the missing terms are contributions of excited states.
The normalization $Z_\pi = \bra{0} P \ket{\pi(\bzero)}$ represents the overlap factor, which also contributes to the pion two-point function. The validity of the spectral decomposition depends on the physical ``extension'' of the local field, $\widehat{\mathring{O}}_n(0;t)$, being smaller than the separation with the pion interpolators. Using $\sqrt{8t}$ as a rough estimate of the size of $\widehat{\mathring{O}}_n(0;t)$, ideally we would work in the region  
where $\sqrt{8t} \ll x_4 - y_4$ and $\sqrt{8t} \ll y_4$. However, one should verify numerically whether the 3-point functions are projected to the ground state at finite flow time. 

We are interested in determining the matrix elements $ \mathring{A}_n(t)$ defined in Eq.~\eqref{eq:sprectral}. Applying the SFTX in Eq.~\eqref{eq:sftx} to $\widehat{\mathring{O}}_n(0;t)$ is then possible, using the perturbative matching coefficients in Eq.~\eqref{eq:cn_1l}, to determine the same matrix element with twist-2 operator renormalized for example in the $\MS$ scheme $A^{\MS}_n(\mu)\equiv \bra{\pi(\bzero)}O_n^{\MS}(\mu)\ket{\pi(\bzero)}$. 

As we have shown in the previous section, with this new method, we have in principle the freedom to choose for the fields $\widehat{\mathring{O}}_n(0;t)$ any Lorentz indices. To avoid the use of external spatial momentum it is convenient to choose all temporal indices, and from now on with $\widehat{\mathring{O}}_n(0;t)$ we denote a field with $n$ indices given by $\mu_1 \cdots \mu_n = 4 \cdots 4$. We emphasize that because we 
want to determine the matrix elements of a traceless operator we need to subtract contributions with operators that include spatial indices (see e.g. Eq.~\eqref{eq:On4}). 
For example, for the cases $n=2,3,4$ a total of $4$, $10$, and $40$ unique index combinations are needed:
\bea
\widehat{\mathring{O}}_{n=2}&=&\mathring{O}_{44}-\frac{1}{3}\sum_{i=1}^3\mathring{O}_{ii}\;, \\
\widehat{\mathring{O}}_{n=3} &=& \mathring{O}_{444}-\frac{1}{3}\sum_{i=1}^3(\mathring{O}_{ii4}+\mathring{O}_{i4i}+\mathring{O}_{4ii})\,, \\
\widehat{\mathring{O}}_{n=4} &=& \mathring{O}_{4444}+\frac{1}{5}\sum_{i=1}^3 \mathring{O}_{iiii}+\frac{1}{15}\sum_{i,j=1,j>i}^3 (\mathring{O}_{iijj}+\mathring{O}_{ijij}+\mathring{O}_{ijji}+\mathring{O}_{jjii}+\mathring{O}_{jiji}+\mathring{O}_{jiij}) \\
&-&\frac{1}{3}\sum_{i=1}^3 (\mathring{O}_{ii44}+\mathring{O}_{i4i4}+\mathring{O}_{i44i}+\mathring{O}_{44ii}+\mathring{O}_{4i4i}+\mathring{O}_{4ii4})\;, 
\label{eq:O4444} 
\eea
where the traceless operators are normalized such that the coefficient of the operator with all temporal Lorentz indices is set to one.
Using Lorentz symmetry and converting to a Euclidean metric the parametrization of the matrix elements is then given by 
\bea
& A^{\MS}_{n=2}(\mu) & =-2 m_\pi^2 \braket{x}^{\MS}(\mu)\,,\\ 
& A^{\MS}_{n=3}(\mu) & =~2 m_\pi^3 \braket{x^2}^{\MS}(\mu)\,,\\
& A^{\MS}_{n=4}(\mu) & =- 2 m_\pi^4 \braket{x^3}^{\MS}(\mu)\,.
\eea 

Although the method allows for the independent calculation of all moments, as a first numerical test, we find it convenient to compute ratios of flowed moments, as in Eq.~\eqref{eq:Rn}. By using the matching coefficients $c_n(t,\mu)$, it is then possible to determine all moments normalized with respect to $\braket{x}_{\MS}$. Since the calculation of $\braket{x}$
is now straightforward and presents no significant difficulties (see Refs.~\cite{Alexandrou:2024zvn,Loffler:2021afv} for recent determinations), we can easily convert the ratio results into values for the individual moments.
The calculation of ratios as in Eq.~\eqref{eq:Rn} is convenient for many reasons: it should improve the statistical fluctuations, it might contain cancellations for cutoff effects and higher order terms in the SFTX, and it could additionally present advantages in terms of the perturbative matching~\cite{Shindler:2023xpd}.
Finally, ratios of flowed moments do not need the calculation of ringed fields with the UV divergences canceling out.
This implies that numerically we calculate ratios of correlation functions~\eqref{eq:3ptfull} with different $n$. 

The correlation function for generic $n$ and generic Lorentz indices $\mu_1\mu_2...\mu_n$ that contributes to Eq.~\eqref{eq:3ptfull} can be written contracting the fermion fields as 
\bea  
C^{3\text{pt}}_{\mu_1...\mu_n}(x_4,y_4;t) &=& -a^6\sum_{\by,\bx} \left\langle \tr\left\{\gamma_5 S_d(0,x) \gamma_5~ a^4 \sum_w S_u(x,w) K(y,w;t,0)^\dagger \Gamma_n(y,t)~\times \right. \right. \nonumber \\ 
&\times & \left. \left. a^4 \sum_v K(y,v;t,0)S_u(v,0)\right\}\right\rangle_{\text{G}}\,,
\label{eq:3pt_flowed_prop}
\eea  
where $S_{u,d}(y,0)$ is the $u$ (or $d$) quark propagator from $0$ to $y$, and the subscript G indicates the gauge average. To compute the remaining contributions one first computes the sequential propagator $\Sigma_{ud}(y,0;t_s) = a^3 \sum_{\bx}S_u(y,x_s) \gamma_5 S_d(x_s,0)$ as solution of the equation 
\be 
a\sum_{v}D_u(x,v)\Sigma_{ud}(v,0;t_s) = \gamma_5 S_d(x,0)\delta_{x_4 t_s}\,,
\label{eq:dirac_seq}
\ee 
where $x_4=t_s$ is the fixed time location of the sink, $x_s=(\bx,t_s)$.
The twist-2 operators for generic $n$ and generic Lorentz indices are denoted with $\Gamma_n(y,t) = \gamma_{\mu_1}\lrDmu2 \cdots \lrD_{\mu_n}$ that indicates where the operator is inserted, $y=(\by,\tau)$, and its flow-time, $t$, dependence.
The flowed propagator 
\be 
S_u(y,0;t) = a^4 \sum_v K(y,v;t,0)S_u(v,0)\,,
\label{eq:flowed_prop}
\ee 
is obtained solving the GF equation with initial condition $S_u(y,0)$. The other contribution to the correlation is obtained solving the GF equation 
\be 
\begin{cases} 
(\partial_t - \Delta_y) \Sigma_{ud}(y,0;t_s;t) = 0 \\
\Sigma_{ud}(y,0;t_s;t=0) = \Sigma_{ud}(y,0;t_s)\,,
\end{cases}
\ee 
where the initial condition is the generalized propagator.
The correlation function~\eqref{eq:3pt_flowed_prop} can then be evaluated with 
\be 
C^{3\text{pt}}_{\mu_1...\mu_n}(t_s,\tau;t) = -a^3\sum_{\by} \left\langle \tr\left\{\Sigma_{ud}^\dagger(0,y;t_s;t) \gamma_5 \Gamma_n(y,t) S_u(y,0;t)\right\}\right\rangle_{\text{G}}\,.
\label{eq:3pt_flowed}
\ee

Eq.~\eqref{eq:3pt_flowed} contains $4^{n-1}$ terms due to the stacked symmetric covariant derivatives, individually defined 
as
\bea 
\phibar(x)\lrD_\mu\phi(x) &=& \frac{1}{2}\phibar(x)\left(\rDmu-\lDmu\right)\phi(x) \\
&=& \frac{1}{4a}\left(\phibar(x)U_{\mu}(x)\phi(x+a\hat{\mu})-\phibar(x)U^{\dagger}_{\mu}(x-a\hat{\mu})\phi(x-a\hat{\mu}) \right.\nonumber\\
&-& \left.\phibar(x+a\hat{\mu})U_{\mu}^{\dagger}(x)\phi(x)+\phibar(x-a\hat{\mu})U_{\mu}(x-a\hat{\mu})\phi(x)\right)\,,
\eea
where to simplify the notation we consider the derivative to act on generic fields $\phi$ and $\phibar$.
This results in a high contraction cost as $n$ increases. To reduce that, we note that the two terms in $\overleftarrow{D}_{\mu}$ (last line) are the same as those in $\overrightarrow{D}_{\mu}$, except the one with the gauge link pointing to the forward $\mu$-direction is shifted by $+\mu$, and the one with the gauge link pointing to the backward $\mu$-direction is shifted by $-\hat{\mu}$. This is of course true for any number of derivatives, i.e., we can always obtain the terms in
\be \label{eq:bwdderiv}
\phibar(x)\lrDmu2 \cdots\lD_{\mu_i} \cdots\lrD_{\mu_n} \phi(x)
\ee
from the terms in
\be \label{eq:fwdderiv}
\phibar(x)\lrDmu2 \cdots \rD_{\mu_i} \cdots \lrD_{\mu_n} \phi(x)
\ee
by appropriately shifting each term. That means that we only need to compute the $2^{n-1}$ terms in
\be
\phibar(x)\rD_{\mu_2} \cdots \rD_{\mu_n} \phi(x)
\ee
and then reconstruct the full
\be
\phibar(x)\lrDmu2 \cdots \lrD_{\mu_n} \phi(x)
\ee
by shifting and linearly combining these terms. 

For the 3-point function~\eqref{eq:3pt_flowed}, this can be written compactly by first defining:
\begin{align}
\mathcal{P}^{1}_{\mu}S_u(x,0;t) &\equiv U_{\mu}(x)S_u(x+a\hat{\mu},0;t) \;, \\
\mathcal{P}^{0}_{\mu}S_u(x,0;t) &\equiv -U^{\dagger}_{\mu}(x-a\hat{\mu})S_u(x-a\hat{\mu},0;t) \;.
\end{align}
\be
\mathcal{C}_{\mu_1\mu_2...\mu_n}(t_s,\tau; t; \ell_2...,\ell_{n}) = -a^3\sum_{\by}\left\langle\text{tr}\left\{ \Sigma^{\dagger}_{ud}(0,y;t_s;t)\gamma_5\gamma_{\mu_1}\mathcal{P}^{\ell_2}_{\mu_2}\mathcal{P}^{\ell_3}_{\mu_3}...\mathcal{P}^{\ell_{n}}_{\mu_{n}}S_{u}(y,0;t)\right\}\right\rangle_{\text{G}} \;,
\ee
where $\ell_i \in\{1,0\}$. Eq.~\eqref{eq:3pt_flowed} can then be written as 
\bea
&& C_{\mu_1...\mu_n}^{3\text{-pt}}(t_s,\tau; t) =   \\ && \frac{1}{2^{n-1}}\sum_{\ell_i\in\{1,0\}}\frac{1}{(2a)^{n_4}}\sum_{k=0}^{n_4}\binom{n_4}{k}\mathcal{C}_{\mu_1\mu_2...\mu_n}\bigg(t_s,\tau-\Delta \tau(\mu_2,...,\mu_n; \ell_2,...,\ell_n)+k;t; \ell_2,\ell_3,...,\ell_n\bigg) \;, \nonumber
\eea
where 
\begin{equation}
\Delta \tau(\mu_2,...,\mu_n; \ell_2,...,\ell_n) = \sum_{i \; \text{with} \; \mu_i=4} \ell_i \;,
\end{equation} 
and $n_4$ is the number of covariant derivatives with temporal indices.

For covariant derivatives with $\mu_i\in\{1,2,3\}$, shifting and linearly combining terms is unnecessary; after projecting into definite 3-momentum, the terms in Eq.~\eqref{eq:bwdderiv} are numerically identical to those in Eq.~\eqref{eq:fwdderiv} up to an overall phase, which is unity in the zero-momentum case considered here. Thus, to reconstruct the full three-point function, it is sufficient to compute only $2^{n_4}$ terms.

Further optimization can be achieved, especially when calculating multiple Mellin moments $n$, by sorting the operators with distinct indices via a depth-first search algorithm. Intermediate results can then be reused effectively—for example, $\mcP_1 S$ can be reused within expressions like $\mcP_1 \mcP_1 S$.

\section{Preliminary results}
\label{sec:results}

\vspace{-1.5ex}
\begin{table*}[h]
\begin{center}
\begin{tabular}{c|cccccc}
\hline
\hline
  label & $a~(\text{fm})$  & $t_0/a^2$ & $\beta$ & $\kappa_{ud}=\kappa_s$ & dimension& N$_{\text{cfg}}$ \\
 \hline
  {\tt a12m412\_mL6.0}  & 0.12 & 1.48554(43) & 3.685& 0.1394305 & $96\times 24^3$ & 119 \\
  {\tt a094m412\_mL6.2} & 0.094 & 2.44003(79) & 3.8  & 0.138963 & $96\times 32^3$ & 210 \\
  {\tt a077m412\_mL7.7} & 0.077 & 3.6200(10) & 3.9 & 0.138603 & $96\times 48^3$ & 200 \\
  {\tt a064m412\_mL6.4} & 0.064 & 5.2425(21) & 4.0  & 0.138272 & $96\times 48^3$ & 200 \\
\end{tabular}
\caption{\label{tab:gauges} Details of the SWF lattice ensembles adopted for this numerical test. All the ensembles have been generated at the $SU(3)$ flavor symmetric point, i.e. all $3$ quark masses are degenerate and have been tuned such that the pseudoscalar mass is $\simeq 411~\text{MeV}$. The listed values of $t_0/a^2$ with corresponding uncertainties are preliminary and have been computed using a larger set of gauge configurations.}
\end{center}
\end{table*}

The 3-point functions are computed as described in the previous section on four $SU(3)$-flavor-symmetric ensembles with Stabilized Wilson Fermions (SWF)~\cite{Francis:2019muy} generated by the OpenLat initiative~\cite{Francis:2022hyr,Cuteri:2022oms,Cuteri:2022erk,Francis:2023gcm} with fixed pseudoscalar mass, $m_{\pi}\simeq 411~\text{MeV}$ and four different lattice spacings ranging between $a\simeq 0.064~\text{fm}$ and $a\simeq 0.12~\text{fm}$. The ensemble parameters, along with the number of configurations used for each, are tabulated in Table~\ref{tab:gauges}. The scale is set using the gradient flow, and converted to $\text{fm}$ via $\sqrt{8t_0}=0.4091(25) ~\text{fm}$~\cite{FlavourLatticeAveragingGroupFLAG:2021npn}. We use one stochastic source per configuration, with source locations randomly sampled from a uniform 4D probability distribution, and a single sequential source at sink time $t_s=40 a$. The measurement is done for flow times up to $t/a^2 = 4.3$ for {\tt a12m412\_mL6.0}, $t/a^2 = 7.0$ for {\tt a094m412\_mL6.2}, $t/a^2 = 10.0$ for {\tt a077m412\_mL7.7}, and $t/a^2 = 11.7$ for {\tt a064m412\_mL6.4}, to cover a range up to approximately $t/t_0\sim 2.5-3.0$ for all ensembles. Reweighting factors are calculated stochastically and  bootstrap resampling with $500$ bootstrap samples is used to determine the statistical uncertainties.

\begin{figure}
    \centering
    \includegraphics[width=0.98\linewidth]{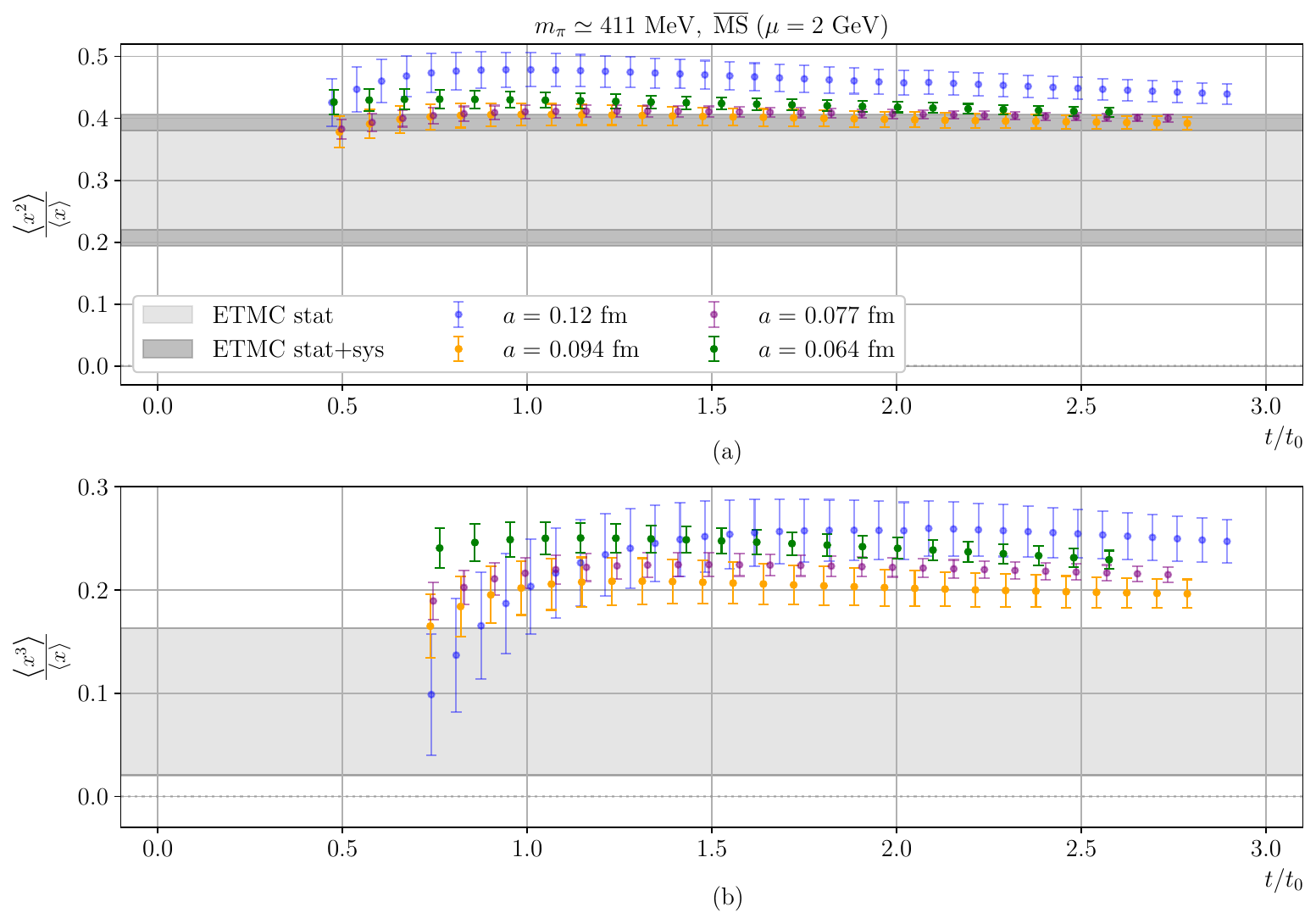}
    \caption{Flow time dependence of ratios of moments of parton distribution functions. Different colors correspond to different lattice spacings. The lattice QCD results are shown in the $\MSbar$ scheme after the NLO matching. For comparison we show results obtained in Ref.~\cite{Alexandrou:2020gxs} (top panel), and in Ref.~\cite{Alexandrou:2021mmi} (middle and bottom panel).}
    \label{fig:ratios}
\end{figure}

In Fig.~\ref{fig:ratios}, we show the two ratios of three-point functions as functions of the flow time in units of
$t_0$
\be
-\frac{c_2(t,\mu)C^{3\text{-pt}}_{n=3}(t)}{m_\pi c_3(t,\mu)C^{3\text{-pt}}_{n=2}(t)}\,, \quad  \frac{c_2(t,\mu)C^{3\text{-pt}}_{n=4}(t)}{m_\pi^2 c_4(t,\mu)C^{3\text{-pt}}_{n=2}(t)}\;,
\ee
with a fixed source-sink separation of $t_s=40a$ and fixed operator insertion time of $\tau=20a$. 
When the ratios of 3-point functions are multiplied by the NLO perturbative matching factors, $c_n(t,\mu=2~\text{GeV})$ from Eq.~\eqref{eq:matching_1l}, and normalized by appropriate powers of $m_\pi$, they correspond respectively to 
\be
\frac{\braket{x^2}}{\braket{x}},\frac{\braket{x^3}}{\braket{x}}\;\;\text{at} \;\;\overline{\text{MS}}, \mu=2~\text{GeV} \;,
\ee 
after taking the continuum limit, up to a residual flow-time dependence from higher-dimensional operators. In the matching coefficients, we have used $\alphas(\mu=2~\text{GeV}) = 0.3069$ determined using the \textit{RunDec} package~\cite{Chetyrkin:2000yt}.
These ratios show a very smooth dependence on the flow time, which is promising for the extrapolation to $t\rightarrow 0$ needed to determine the moments in the $\MSbar$ scheme. As expected, we observe more pronounced cutoff effects at shorter flow times, with these effects appearing at smaller $t/t_0$ for the finer ensembles. For lattice spacings $a \lesssim 0.094$ fm, the cutoff effects remain minimal over a wide range of $t/t_0$.
Preliminary observations suggest that, at fixed lattice spacing, the range where short-distance discretization errors 
impact the flow time dependence shifts to larger $t/t_0$
values for higher moments.
This is likely due to the larger time extent, in lattice units, of the twist-2 local operators.

In Fig.~\ref{fig:ratios}, we compare  against calculations of the same quantities presented in Refs.~\cite{Alexandrou:2020gxs,Alexandrou:2021mmi}, which were obtained with $N_f=2+1+1$ flavors of twisted-mass fermions and a lighter pion mass of $m_{\pi}\simeq 260~\text{MeV}$. These results were obtained with the canonical approach of choosing irreducible representations of the hypercubic group H($4$) for which boosting the 3-point functions in one or more directions is necessary in order to avoid mixing with lower-dimensional operators. Due to the noise introduced from boosting, this requires a much larger number of measurements; indeed, the results we compare against used $N_{\text{cfg}}\times N_{\text{source}} = 3904$ for $\braket{x^2}$~\cite{Alexandrou:2020gxs} and $N_{\text{cfg}}\times N_{\text{source}}\times N_{\text{boost}}= 15-70\text{k}$ for $\braket{x^3}$~\cite{Alexandrou:2021mmi}. 
The method studied in this work achieves over twice the precision with only $100-200$ measurements, thanks to the favorable noise properties of the unboosted three-point functions and the suppression of ultraviolet fluctuations via the gradient flow.
Preliminary results suggest that, assuming an equivalent number of independent measurements, statistical errors with the new method could be reduced by a factor of 
$\sim 10$ for $\braket{x^2}$ and by a factor $\sim 40$ for $\braket{x^3}$.
Any tension with the older results may stem from the different quark masses used. These will be investigated through a comprehensive analysis of the three-point functions at various sink and operator insertion times, along with an assessment of the residual flow time dependence of the ratios, in our forthcoming publication.

\begin{figure}
    \centering
    \includegraphics[width=0.98\linewidth]{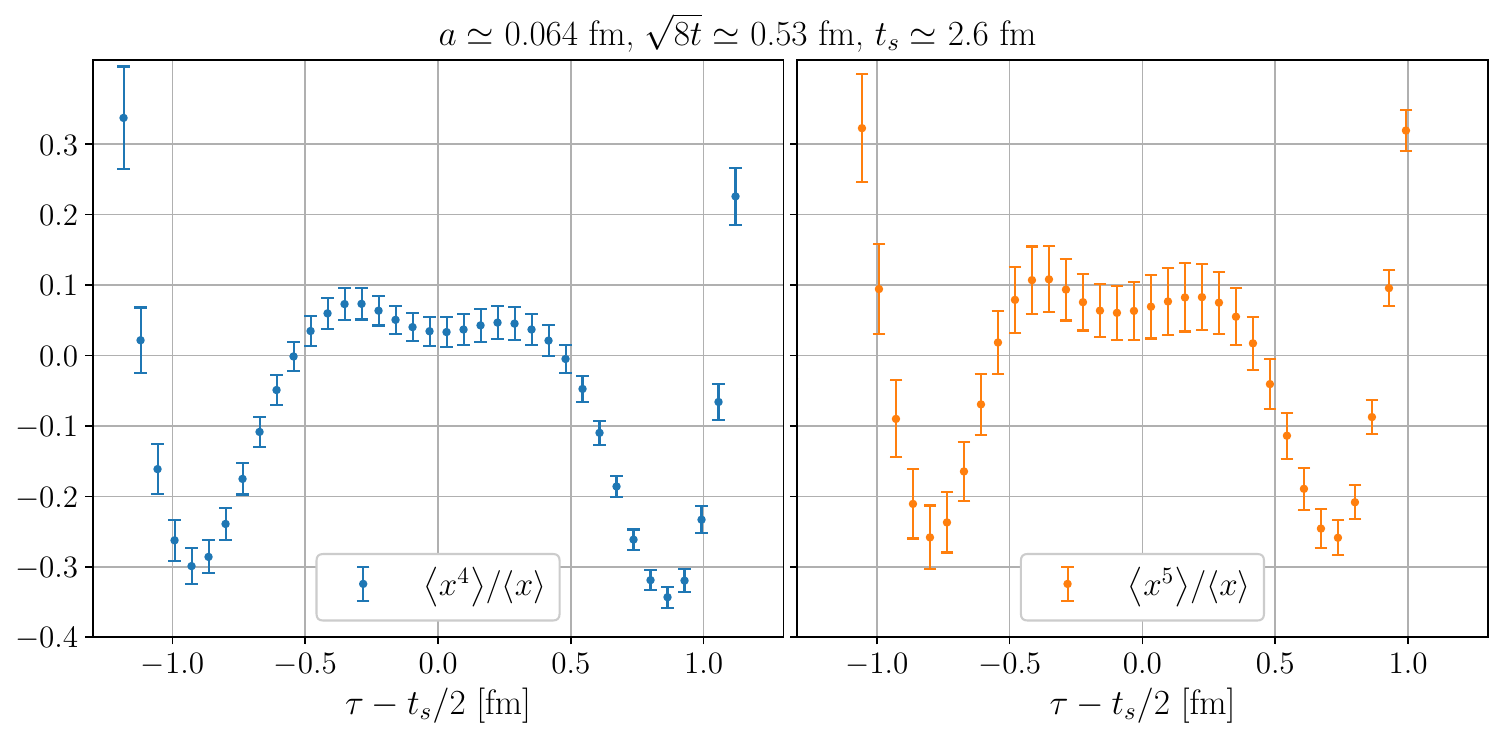}
    \caption{Euclidean time dependence of ratios of  correlation functions in Eq.~\eqref{eq:3ptfull} corresponding to $\left\langle x^4 \right\rangle/\left\langle x \right\rangle$ and $\left\langle x^5 \right\rangle/\left\langle x \right\rangle$ computed at a fixed flow time using ensemble {\tt a064m412\_mL6.4}. The numerical results are matched to represent the ratios in the $\MSbar$ scheme at scale $\mu=2$ GeV.}
    \label{fig:plateaus}
\end{figure}

The new method resolves also the problem of power divergences. In Fig.~\ref{fig:plateaus} we show examples of the Euclidean time dependence of the correlation functions in Eq.~\eqref{eq:3ptfull} corresponding to the ratios $\left\langle x^4 \right\rangle/\left\langle x \right\rangle$ and $\left\langle x^5 \right\rangle/\left\langle x \right\rangle$. 
These results, albeit preliminary, are very encouraging showing that, at least in the pseudoscalar channel, the method promises a precise reconstruction of the full PDF.

\section{Conclusion}
\label{sec:conclusion}

We have performed the first numerical investigation of the recently proposed method~\cite{Shindler:2023xpd} to obtain precise ratios of arbitrary-order moments of PDFs. We present preliminary results for the ratios $\braket{x^2}/\braket{x}$ and $\braket{x^3}/\braket{x}$ in the $\MSbar$ scheme at $\mu=2$ GeV. As a testing ground we have considered PDFs of the pseudoscalar meson using four OpenLat ensembles with $m_{\pi}\simeq 411~\text{MeV}$. We find that this method achieves significantly higher precision, requiring approximately $20$ times fewer measurements for $\braket{x^2}/\braket{x}$ and about $350$ times fewer measurements for $\braket{x^3}/\braket{x}$ compared to the current state-of-the-art lattice QCD results for these quantities. We have also shown that the method allows to solve the problem of the power divergences and allows to calculate moments up to $\left\langle x^5 \right\rangle$ with a very moderate computational effort. 
These preliminary results seem to give first indication that the continuum and $ t \rightarrow 0$ limits should be well under control. 
Additional systematics can arise from excited state contamination.
All these questions will be addressed in the upcoming publication where we
will increase our statistics, and compute additional source-sink separations for all
the ensembles. After performing the continuum limit at fixed flow time in
physical units we will extrapolate the residual flow time dependence. 
We have shown first results also for higher moments such as $\braket{x^4}$ and $\braket{x^5}$ and we plan to finalize the analysis to attempt the reconstruction of the PDF based on these moments.

\section*{Acknowledgments}
\noindent 
Numerical calculations for this work were performed using resources from the National Energy Research Scientific Computing Center (NERSC), a Department of Energy Office of Science User Facility, under NERSC award NP-ERCAP0027662; and the Gauss Centre for Supercomputing e.V. (www.gauss-centre.eu) on the GCS Supercomputer JUWELS~\cite{juelich2021juwels} at the Jülich Supercomputing Centre (JSC).
We also acknowledge the EuroHPC Joint Undertaking for awarding this project access to the EuroHPC supercomputer LEONARDO, hosted by CINECA (Italy) and LUMI at CSC (Finland).
The authors acknowledge support as well as computing and storage resources by GENCI on Adastra and Occigen (CINES), Jean-Zay (IDRIS) and Ir\`ene-Joliot-Curie (TGCC) under projects (2020-2024)-A0080511504 and (2020-2024)-A0080502271.

A.F. acknowledges support by the National Science and Technology Council of Taiwan under grant 111-2112-M-A49-018-MY2.
The work of R.K. is supported in part by the NSF Graduate Research Fellowship Program under Grant DGE-2146752. 
J.K. is supported by the Deutsche Forschungsgemeinschaft (DFG) through the CRC-TR 211 'Strong-interaction matter under extreme conditions'– project number 315477589 – TRR211.
G.P. acknowledges funding by the Deutsche Forschungsgemeinschaft (DFG, German Research Foundation) - project number 460248186 (PUNCH4NFDI).
D.A.P is supported from the Office of Nuclear Physics, Department of Energy, under contract DE-SC0004658. 
A.S. acknowledges funding support from Deutsche Forschungsgemeinschaft (DFG, German Research Foundation) through grant 513989149 and under the National Science Foundation grant PHY-2209185. The work of A.W-L. was supported by the U.S. Department of Energy, Office of Science, Office of Nuclear Physics, under contract number DE-AC02-05CH11231.
The research of S.Z. is funded, in part, by l’Agence Nationale de la Recherche (ANR), project ANR-23-CE31-0019. For the purpose of open access, the author has applied a CC-BY public copyright licence to any Author Accepted Manuscript (AAM) version arising from this submission.
We acknowledge support from the DOE Topical Collaboration “Nuclear Theory for New Physics”, award No. DE-SC0023663.

The {\tt Chroma}~\cite{Edwards:2004sx},  {\tt QUDA}~\cite{Clark:2009wm,Babich:2011np,Clark:2016rdz}, {\tt QDP-JIT}~\cite{6877336}, and {\tt LALIBE}~\cite{lalibe} software libraries were used in this work. 
Data analysis used {\tt NumPy}~\cite{harris2020array}, {\tt SciPy}~\cite{2020SciPy-NMeth}, and {\tt gvar}~\cite{peter_lepage_2020_4290884}.
Figures were produced using {\tt matplotlib}~\cite{Hunter:2007}.

\bibliographystyle{JHEP}
\bibliography{flowed_mom}

\providecommand{\href}[2]{#2}\begingroup\raggedright\begin{thebibliography}{10}

\bibitem{Shindler:2023xpd}
A.~Shindler, \emph{{Moments of parton distribution functions of any order from
  lattice QCD}},
  \href{https://doi.org/10.1103/PhysRevD.110.L051503}{\emph{Phys. Rev. D}
  {\bfseries 110} (2024) L051503}
  [\href{https://arxiv.org/abs/2311.18704}{{\ttfamily 2311.18704}}].

\bibitem{Cichy:2018mum}
K.~Cichy and M.~Constantinou, \emph{{A guide to light-cone PDFs from Lattice
  QCD: an overview of approaches, techniques and results}},
  \href{https://doi.org/10.1155/2019/3036904}{\emph{Adv. High Energy Phys.}
  {\bfseries 2019} (2019) 3036904}
  [\href{https://arxiv.org/abs/1811.07248}{{\ttfamily 1811.07248}}].

\bibitem{Karpie:2018zaz}
J.~Karpie, K.~Orginos and S.~Zafeiropoulos, \emph{{Moments of Ioffe time parton
  distribution functions from non-local matrix elements}},
  \href{https://doi.org/10.1007/JHEP11(2018)178}{\emph{JHEP} {\bfseries 11}
  (2018) 178} [\href{https://arxiv.org/abs/1807.10933}{{\ttfamily
  1807.10933}}].

\bibitem{Joo:2019bzr}
B.~Jo\'o, J.~Karpie, K.~Orginos, A.V.~Radyushkin, D.G.~Richards, R.S.~Sufian
  et~al., \emph{{Pion valence structure from Ioffe-time parton
  pseudodistribution functions}},
  \href{https://doi.org/10.1103/PhysRevD.100.114512}{\emph{Phys. Rev. D}
  {\bfseries 100} (2019) 114512}
  [\href{https://arxiv.org/abs/1909.08517}{{\ttfamily 1909.08517}}].

\bibitem{Gao:2022iex}
X.~Gao, A.D.~Hanlon, N.~Karthik, S.~Mukherjee, P.~Petreczky, P.~Scior et~al.,
  \emph{{Continuum-extrapolated NNLO valence PDF of the pion at the physical
  point}}, \href{https://doi.org/10.1103/PhysRevD.106.114510}{\emph{Phys. Rev.
  D} {\bfseries 106} (2022) 114510}
  [\href{https://arxiv.org/abs/2208.02297}{{\ttfamily 2208.02297}}].

\bibitem{Orginos:2017kos}
K.~Orginos, A.~Radyushkin, J.~Karpie and S.~Zafeiropoulos, \emph{{Lattice QCD
  exploration of parton pseudo-distribution functions}},
  \href{https://doi.org/10.1103/PhysRevD.96.094503}{\emph{Phys. Rev. D}
  {\bfseries 96} (2017) 094503}
  [\href{https://arxiv.org/abs/1706.05373}{{\ttfamily 1706.05373}}].

\bibitem{Kronfeld:1984zv}
A.S.~Kronfeld and D.M.~Photiadis, \emph{{Phenomenology on the Lattice:
  Composite Operators in Lattice Gauge Theory}},
  \href{https://doi.org/10.1103/PhysRevD.31.2939}{\emph{Phys. Rev. D}
  {\bfseries 31} (1985) 2939}.

\bibitem{Martinelli:1987zd}
G.~Martinelli and C.T.~Sachrajda, \emph{{Pion Structure Functions From Lattice
  {QCD}}}, \href{https://doi.org/10.1016/0370-2693(87)90601-0}{\emph{Phys.
  Lett. B} {\bfseries 196} (1987) 184}.

\bibitem{Martinelli:1987bh}
G.~Martinelli and C.T.~Sachrajda, \emph{{A Lattice Calculation of the Pion's
  Form-Factor and Structure Function}},
  \href{https://doi.org/10.1016/0550-3213(88)90445-2}{\emph{Nucl. Phys. B}
  {\bfseries 306} (1988) 865}.

\bibitem{Narayanan:2006rf}
R.~Narayanan and H.~Neuberger, \emph{{Infinite N phase transitions in continuum
  Wilson loop operators}},
  \href{https://doi.org/10.1088/1126-6708/2006/03/064}{\emph{JHEP} {\bfseries
  0603} (2006) 064} [\href{https://arxiv.org/abs/hep-th/0601210}{{\ttfamily
  hep-th/0601210}}].

\bibitem{Luscher:2010iy}
{L\"uscher, M.}, \emph{{Properties and uses of the Wilson flow in lattice
  QCD}}, \href{https://doi.org/10.1007/JHEP08(2010)071}{\emph{JHEP} {\bfseries
  1008} (2010) 071} [\href{https://arxiv.org/abs/1006.4518}{{\ttfamily
  1006.4518}}].

\bibitem{Luscher:2011bx}
{L\"uscher, M.} and P.~Weisz, \emph{{Perturbative analysis of the gradient flow
  in non-abelian gauge theories}},
  \href{https://doi.org/10.1007/JHEP02(2011)051}{\emph{JHEP} {\bfseries 1102}
  (2011) 051} [\href{https://arxiv.org/abs/1101.0963}{{\ttfamily 1101.0963}}].

\bibitem{Luscher:2013cpa}
{L\"uscher, M.}, \emph{{Chiral symmetry and the Yang--Mills gradient flow}},
  \href{https://doi.org/10.1007/JHEP04(2013)123}{\emph{JHEP} {\bfseries 1304}
  (2013) 123} [\href{https://arxiv.org/abs/1302.5246}{{\ttfamily 1302.5246}}].

\bibitem{Makino:2014taa}
H.~Makino and H.~Suzuki, \emph{{Lattice energy--momentum tensor from the
  Yang-Mills gradient flow--inclusion of fermion fields}},
  \href{https://doi.org/10.1093/ptep/ptu070}{\emph{PTEP} {\bfseries 2014}
  (2014) 063B02} [\href{https://arxiv.org/abs/1403.4772}{{\ttfamily
  1403.4772}}].

\bibitem{Luscher:2013vga}
M.~Lüscher, \emph{{Future applications of the Yang-Mills gradient flow in
  lattice QCD}}, \href{https://doi.org/10.22323/1.187.0016}{\emph{PoS}
  {\bfseries LATTICE2013} (2014) 016}
  [\href{https://arxiv.org/abs/1308.5598}{{\ttfamily 1308.5598}}].

\bibitem{Manohar:2018aog}
A.V.~Manohar, \emph{{Introduction to Effective Field Theories}},
  \href{https://arxiv.org/abs/1804.05863}{{\ttfamily 1804.05863}}.

\bibitem{Mereghetti:2021nkt}
E.~Mereghetti, C.J.~Monahan, M.D.~Rizik, A.~Shindler and P.~Stoffer,
  \emph{{One-loop matching for quark dipole operators in a gradient-flow
  scheme}}, \href{https://doi.org/10.1007/JHEP04(2022)050}{\emph{JHEP}
  {\bfseries 04} (2022) 050}
  [\href{https://arxiv.org/abs/2111.11449}{{\ttfamily 2111.11449}}].

\bibitem{Crosas:2023anw}
O.L.~Crosas, C.J.~Monahan, M.D.~Rizik, A.~Shindler and P.~Stoffer,
  \emph{{One-loop matching of the CP-odd three-gluon operator to the gradient
  flow}}, \href{https://doi.org/10.1016/j.physletb.2023.138301}{\emph{Phys.
  Lett. B} {\bfseries 847} (2023) 138301}
  [\href{https://arxiv.org/abs/2308.16221}{{\ttfamily 2308.16221}}].

\bibitem{Gross:1974cs}
D.J.~Gross and F.~Wilczek, \emph{{Asymptotically free gauge theories 2.}},
  \href{https://doi.org/10.1103/PhysRevD.9.980}{\emph{Phys. Rev. D} {\bfseries
  9} (1974) 980}.

\bibitem{Ablinger:2013cf}
J.~Ablinger, J.~Bl\"umlein and C.~Schneider, \emph{{Analytic and Algorithmic
  Aspects of Generalized Harmonic Sums and Polylogarithms}},
  \href{https://doi.org/10.1063/1.4811117}{\emph{J. Math. Phys.} {\bfseries 54}
  (2013) 082301} [\href{https://arxiv.org/abs/1302.0378}{{\ttfamily
  1302.0378}}].

\bibitem{Artz:2019bpr}
J.~Artz, R.V.~Harlander, F.~Lange, T.~Neumann and M.~Prausa, \emph{{Results and
  techniques for higher order calculations within the gradient-flow
  formalism}}, \href{https://doi.org/10.1007/JHEP06(2019)121}{\emph{JHEP}
  {\bfseries 06} (2019) 121}
  [\href{https://arxiv.org/abs/1905.00882}{{\ttfamily 1905.00882}}].

\bibitem{Kim:2021qae}
{\scshape SymLat} collaboration, \emph{{Nonperturbative renormalization of the
  quark chromoelectric dipole moment with the gradient flow: Power
  divergences}}, \href{https://doi.org/10.1103/PhysRevD.104.074516}{\emph{Phys.
  Rev. D} {\bfseries 104} (2021) 074516}
  [\href{https://arxiv.org/abs/2106.07633}{{\ttfamily 2106.07633}}].

\bibitem{Cuteri:2022erk}
F.~Cuteri, A.~Francis, P.~Fritzsch, G.~Pederiva, A.~Rago, A.~Shindler et~al.,
  \emph{{Benchmark Continuum Limit Results for Spectroscopy with Stabilized
  Wilson Fermions}}, \href{https://doi.org/10.22323/1.430.0074}{\emph{PoS}
  {\bfseries LATTICE2022} (2023) 074}
  [\href{https://arxiv.org/abs/2212.11048}{{\ttfamily 2212.11048}}].

\bibitem{Cuteri:2022oms}
F.~Cuteri, A.S.~Francis, P.~Fritzsch, G.~Pederiva, A.~Rago, A.~Shindler et~al.,
  \emph{{Gauge generation and dissemination in OpenLat}},
  \href{https://doi.org/10.22323/1.430.0426}{\emph{PoS} {\bfseries LATTICE2022}
  (2023) 426} [\href{https://arxiv.org/abs/2212.07314}{{\ttfamily
  2212.07314}}].

\bibitem{Alexandrou:2024zvn}
C.~Alexandrou et~al., \emph{{Quark and gluon momentum fractions in the pion and
  in the kaon}},  \href{https://arxiv.org/abs/2405.08529}{{\ttfamily
  2405.08529}}.

\bibitem{Loffler:2021afv}
{\scshape RQCD} collaboration, \emph{{Mellin moments of spin dependent and
  independent PDFs of the pion and rho meson}},
  \href{https://doi.org/10.1103/PhysRevD.105.014505}{\emph{Phys. Rev. D}
  {\bfseries 105} (2022) 014505}
  [\href{https://arxiv.org/abs/2108.07544}{{\ttfamily 2108.07544}}].

\bibitem{Francis:2019muy}
A.~Francis, P.~Fritzsch, M.~L\"uscher and A.~Rago, \emph{{Master-field
  simulations of O($a$)-improved lattice QCD: Algorithms, stability and
  exactness}}, \href{https://doi.org/10.1016/j.cpc.2020.107355}{\emph{Comput.
  Phys. Commun.} {\bfseries 255} (2020) 107355}
  [\href{https://arxiv.org/abs/1911.04533}{{\ttfamily 1911.04533}}].

\bibitem{Francis:2022hyr}
A.S.~Francis, F.~Cuteri, P.~Fritzsch, G.~Pederiva, A.~Rago, A.~Shindler et~al.,
  \emph{{Properties, ensembles and hadron spectra with Stabilised Wilson
  Fermions}}, \href{https://doi.org/10.22323/1.396.0118}{\emph{PoS} {\bfseries
  LATTICE2021} (2022) 118} [\href{https://arxiv.org/abs/2201.03874}{{\ttfamily
  2201.03874}}].

\bibitem{Francis:2023gcm}
A.~Francis, F.~Cuteri, P.~Fritzsch, G.~Pederiva, A.~Rago, A.~Shindler et~al.,
  \emph{{Progress in generating gauge ensembles with Stabilized Wilson
  Fermions}}, \href{https://doi.org/10.22323/1.453.0048}{\emph{PoS} {\bfseries
  LATTICE2023} (2024) 048} [\href{https://arxiv.org/abs/2312.11298}{{\ttfamily
  2312.11298}}].

\bibitem{FlavourLatticeAveragingGroupFLAG:2021npn}
{\scshape Flavour Lattice Averaging Group (FLAG)} collaboration, \emph{{FLAG
  Review 2021}},
  \href{https://doi.org/10.1140/epjc/s10052-022-10536-1}{\emph{Eur. Phys. J. C}
  {\bfseries 82} (2022) 869}
  [\href{https://arxiv.org/abs/2111.09849}{{\ttfamily 2111.09849}}].

\bibitem{Chetyrkin:2000yt}
K.G.~Chetyrkin, J.H.~Kuhn and M.~Steinhauser, \emph{{RunDec: A Mathematica
  package for running and decoupling of the strong coupling and quark masses}},
  \href{https://doi.org/10.1016/S0010-4655(00)00155-7}{\emph{Comput. Phys.
  Commun.} {\bfseries 133} (2000) 43}
  [\href{https://arxiv.org/abs/hep-ph/0004189}{{\ttfamily hep-ph/0004189}}].

\bibitem{Alexandrou:2020gxs}
{\scshape ETM} collaboration, \emph{{Mellin moments $\langle x \rangle$ and
  $\langle x^2 \rangle$ for the pion and kaon from lattice QCD}},
  \href{https://doi.org/10.1103/PhysRevD.103.014508}{\emph{Phys. Rev. D}
  {\bfseries 103} (2021) 014508}
  [\href{https://arxiv.org/abs/2010.03495}{{\ttfamily 2010.03495}}].

\bibitem{Alexandrou:2021mmi}
{\scshape ETM} collaboration, \emph{{Pion and kaon \ensuremath{\langle
  x^3}\ensuremath{\rangle} from lattice QCD and PDF reconstruction from Mellin
  moments}}, \href{https://doi.org/10.1103/PhysRevD.104.054504}{\emph{Phys.
  Rev. D} {\bfseries 104} (2021) 054504}
  [\href{https://arxiv.org/abs/2104.02247}{{\ttfamily 2104.02247}}].

\bibitem{juelich2021juwels}
{Jülich Supercomputing Centre}, \emph{Juwels cluster and booster: Exascale
  pathfinder with modular supercomputing architecture at juelich supercomputing
  centre}, \href{https://doi.org/10.17815/jlsrf-7-183}{\emph{Journal of
  Large-Scale Research Facilities} {\bfseries 7} (2021) A183}.

\bibitem{Edwards:2004sx}
{\scshape SciDAC Collaboration, LHPC Collaboration, UKQCD Collaboration}
  collaboration, \emph{{The Chroma software system for lattice QCD}},
  \href{https://doi.org/10.1016/j.nuclphysbps.2004.11.254}{\emph{Nucl.Phys.Proc.Suppl.}
  {\bfseries 140} (2005) 832}
  [\href{https://arxiv.org/abs/hep-lat/0409003}{{\ttfamily hep-lat/0409003}}].

\bibitem{Clark:2009wm}
M.~Clark, R.~Babich, K.~Barros, R.~Brower and C.~Rebbi, \emph{{Solving Lattice
  QCD systems of equations using mixed precision solvers on GPUs}},
  \href{https://doi.org/10.1016/j.cpc.2010.05.002}{\emph{Comput. Phys. Commun.}
  {\bfseries 181} (2010) 1517}
  [\href{https://arxiv.org/abs/0911.3191}{{\ttfamily 0911.3191}}].

\bibitem{Babich:2011np}
R.~Babich, M.~Clark, B.~Joó, G.~Shi, R.~Brower and S.~Gottlieb, \emph{{Scaling
  Lattice QCD beyond 100 GPUs}},  in \emph{{SC11 International Conference for
  High Performance Computing, Networking, Storage and Analysis}}, 9, 2011,
  \href{https://doi.org/10.1145/2063384.2063478}{DOI}
  [\href{https://arxiv.org/abs/1109.2935}{{\ttfamily 1109.2935}}].

\bibitem{Clark:2016rdz}
M.A.~Clark, B.~Joó, A.~Strelchenko, M.~Cheng, A.~Gambhir and R.~Brower,
  \emph{{Accelerating Lattice QCD Multigrid on GPUs Using Fine-Grained
  Parallelization}},  \href{https://arxiv.org/abs/1612.07873}{{\ttfamily
  1612.07873}}.

\bibitem{6877336}
F.T.~{Winter}, M.A.~{Clark}, R.G.~{Edwards} and B.~Joó.

\bibitem{lalibe}
A.~Gambhir, D.~Brantley, J.~Chang, B.~Hörz, H.~Monge-Camacho, P.~Vranas
  et~al., ``lalibe.'' \url{https://github.com/callat-qcd/lalibe}, 2018.

\bibitem{harris2020array}
C.R.~Harris, K.J.~Millman, S.J.~van~der Walt, R.~Gommers, P.~Virtanen,
  D.~Cournapeau et~al., \emph{Array programming with {NumPy}},
  \href{https://doi.org/10.1038/s41586-020-2649-2}{\emph{Nature} {\bfseries
  585} (2020) 357}.

\bibitem{2020SciPy-NMeth}
P.~Virtanen, R.~Gommers, T.E.~Oliphant, M.~Haberland, T.~Reddy, D.~Cournapeau
  et~al., \emph{{{SciPy} 1.0: Fundamental Algorithms for Scientific Computing
  in Python}}, \href{https://doi.org/10.1038/s41592-019-0686-2}{\emph{Nature
  Methods} {\bfseries 17} (2020) 261}.

\bibitem{peter_lepage_2020_4290884}
G.P.~Lepage, \emph{{gvar v. 11.9.1}},
  \href{https://arxiv.org/abs/https://github.com/gplepage/gvar}{{\ttfamily
  https://github.com/gplepage/gvar}}.

\bibitem{Hunter:2007}
J.D.~Hunter, \emph{Matplotlib: A 2d graphics environment},
  \href{https://doi.org/10.1109/MCSE.2007.55}{\emph{Computing in Science \&
  Engineering} {\bfseries 9} (2007) 90}.

\end{thebibliography}\endgroup

\end{document}